\documentstyle[sprocl,epsfig]{article}

\newcommand{\sll}{{\tilde{l}}}
\newcommand{\slR}{{\tilde{l}_{\rm R}}}
\newcommand{\slL}{{\tilde{l}_{\rm L}}}
\newcommand{\msl}[1]{m_{\tilde{l_{#1}}}}

\newcommand{\smuR}{{\tilde{\mu}_{\rm R}}}
\newcommand{\smuL}{{\tilde{\mu}_{\rm L}}}
\newcommand{\se}{{\tilde{e}}}
\newcommand{\seR}{{\tilde{e}_{\rm R}}}
\newcommand{\seL}{{\tilde{e}_{\rm L}}}
\newcommand{\st}{{\tilde{\tau}}}
\newcommand{\stR}{{\tilde{\tau}_1}}

\newcommand{\sn}{{\tilde{\nu}}}
\newcommand{\sne}{{\tilde{\nu}_e}}
\newcommand{\snl}{{\tilde{\nu}_l}}

\newcommand{\mseR}{m_{\tilde{e}_{\rm R}}}

\newcommand{\cha}{\tilde{\chi}}
\newcommand{\neu}{\tilde{\chi}^0}

\newcommand{\sF}{{\tilde{f}}}
\newcommand{\msf}[1]{m_{\rm \tilde{f}_{#1}}}

\def\mathswitch#1{\relax\ifmmode#1\else$#1$\fi}
\def\mathswitchr#1{\relax\ifmmode{\mathrm{#1}}\else$\mathrm{#1}$\fi}

\newcommand{\gesim}{\,\raisebox{-.3ex}{$_{\textstyle >}\atop^{\textstyle\sim}$}\,}

\newcommand{\drbar}{{\mathswitch {\overline{\rm DR}}} }
\newcommand{\OO}{{\mathcal O}}

\begin{document}
\begin{flushright}
DESY 04--155\\
FERMILAB--Conf--04--211--T\\[1.5em]
\end{flushright}

\def\thefootnote{\fnsymbol{footnote}}
\title{SLEPTONS: MASSES, MIXINGS, COUPLINGS\ 
\footnote{Presentation given by A.F. at the International Conference on Linear
Colliders \mbox{(LCWS 04)}, Paris, France (April 19--23, 2004); to appear in the
proceedings.}}

\author{A.~FREITAS$^a$, H.-U. MARTYN$^b$, U. NAUENBERG$^c$, P.~M.~ZERWAS$^d$}

\address{$^a$ Fermi National Accelerator Laboratory, Batavia, IL 60510-500,USA\\
	$^b$ I. Physikalisches Institut, RWTH Aachen, D-52074 Aachen, Germany\\
	$^c$ University of Colorado, Boulder, CO 80301, USA\\
	$^d$ Deutsches Elektronen-Synchrotron DESY, D--22603 Hamburg, Germany}


\maketitle\abstracts{
           This presentation summarizes recent work of different groups on 
           the
           analysis of slepton parameters at a TeV linear collider. In particular,
           measurements of the masses, mixings and Yukawa couplings for the
           neutral and charged sleptons of the first/second generation 
           and for the charged slepton sector of the third generation 
           are reviewed. For all relevant
           processes, threshold corrections and higher order corrections 
           in the continuum are available, thus allowing
           high-precision analyses of the parameters in the slepton sector.
}

\def\thefootnote{\alph{footnote}}

\section{Overview}

The analysis of the supersymmetric particle sector at future colliders 
entails a
diverse and challenging experimental program for the measurement of masses,
mixings and couplings. It is imperative to scrutinize
accurately the fundamental symmetry relations of supersymmetry (SUSY), which
are expressed in the identity of gauge couplings ($g$) and the SUSY Yukawa
couplings ($\hat{g}$) between fermions, sfermions and gauginos. In addition,
the precise determination of SUSY breaking parameters from sparticle mass and
mixing measurements establishes the basis for reconstructing the fundamental
breaking mechanism at possibly very high energy scales \cite{gpz}.

The SUSY partners of the leptons, sleptons, can best be studied at a
future high-energy $e^\pm e^-$ linear collider. The masses of the sleptons can
be extracted from measurements of the energy distributions of their decay
products or from scans of the cross-sections at the
pair production threshold. From the analysis of the production cross-sections
in the continuum, the couplings and mixings of the sleptons can be determined.
In this report, recent advances in understanding the physics of sleptons
will be summarized, with particular focus on new developments for the neutral
sleptons (sneutrinos) and the third generation sleptons (staus). For a review of
earlier slepton studies, see {\it e.g.} Ref.~\cite{lcws02}.

In many SUSY breaking scenarios the sleptons are relatively light, leading to
simple decay signatures involving light neutralino/chargino states and leptons.
Tab.~\ref{tab:sps1} summarizes the most important decay modes in the SPS1a
scenario \cite{sps}.%
\renewcommand{\arraystretch}{1.3}%
\begin{table}[tb]
\begin{tabular}[t]{|l||l|l|l@{$\;\;$}l@{$\quad\;\;$}l@{$\;\;$}l@{$\quad\;\;$}l@{$\;\;$}l|}
\hline
 & Mass & Width & \multicolumn{6}{l|}{Decay modes} \\
\hline
$\slR^\pm = \seR^\pm/\smuR^\pm$ & 142.7 & 0.21 & 
				$l^\pm \, \neu_1$ & 100\% &&&& \\
$\slL^\pm = \seL^\pm/\smuL^\pm$ & 202.3 & 0.25 & $l^\pm \, \neu_1$ & 48\% &
				$l^\pm \, \neu_2$ & 19\% &
				$\nu_l \, \cha_1^\pm$ & 33\% \\
\hline
$\st_1^\pm$ & 133.2 & 0.20 & 
				$\tau^\pm \, \neu_1$ & 100\% &&&& \\
$\st_2^\pm$ & 202.3 & 0.34 & $\tau^\pm \, \neu_1$ & 53\% &
				$\tau^\pm \, \neu_2$ & 17\% &
				$\nu_\tau \, \cha_1^\pm$ & 30\% \\
\hline
$\snl = \sne/\sn_\mu$ & 186.0 & 0.16 & $\nu_l \, \neu_1$ & 87\% &
			$\nu_l \, \neu_2$ & 4\% &
			$l^- \, \cha^+_1$ & 10\% \\
\hline
$\sn_\tau$ & 185.1 & 0.15 & $\nu_\tau \, \neu_1$ & 89\% &
			$\nu_\tau \, \neu_2$ & 3\% &
			$\tau^- \, \cha^+_1$ & 8\% \\
\hline \hline
$\neu_1$ & 96.2 & --- & \multicolumn{2}{c}{---} &&&& \\
$\neu_2$ & 176.6 & 0.020 & $\seR^\pm \, e^\mp$ & 6\% &
			$\smuR^\pm \, \mu^\mp$ & 6\% &
			$\stR^\pm \, \tau^\mp$ & 88\% \\
\hline
$\cha_1^\pm$ & 176.1 & 0.014 & $\stR^+ \, \nu_\tau$ & 100\% &&&& \\
\hline
\end{tabular}
\caption{Masses, widths (in GeV) and main decay branching ratios 
of sleptons and of 
light neutralino and chargino states in their decay chain
for the reference point SPS1a.}
\label{tab:sps1}
\end{table}
\renewcommand{\arraystretch}{1}%
The production of sleptons of the second and third generation in $e^+e^-$
collisions proceeds through
s-channel photon and $Z$-boson exchanges
in P-waves with a characteristic rise of the excitation curve 
$\propto \beta^3$ with
$\beta = (1-4\msl{}^2/s)^{1/2}$. 
The production of selectrons \cite{lcws02}
and electron-sneutrinos proceeds in
addition through t-channel neutralino or chargino exchange, respectively.
Selectrons can also be produced in $e^-e^-$ collisions.
Due to the Majorana nature of the neutralinos, some selectron channels are
generated in S-waves, with a steep threshold excitation $\propto \beta$, namely
for $\seR^\pm \seL^\mp$ pairs in $e^+e^-$ annihilation and $\seR^-\seR^-,
\seL^-\seL^-$ pairs in $e^-e^-$ scattering.

\section{Theoretical Picture}


The steep rise of the slepton excitation curves allows the very precise
determination of the slepton masses (see next section). It is
necessary therefore to include effects beyond leading order in  the theoretical
prediction \cite{thr1,slep}. Non-zero width effects play an important role
near threshold and can be incorporated by shifting the slepton mass to the
complex plane, $\msf{}^2 \to \msf{}^2 - i \msf{} \Gamma_{\rm\!\sF}$. 
To keep the
amplitude gauge-invariant, the signal contribution with two resonant sleptons,
$ee \to \tilde{f}\tilde{f} \to ff\cha\cha$, $f = e,\mu,\tau,\nu$, must be
supplemented by other non-resonant diagrams leading to the same final state 
$ff\cha\cha$.
The production 
cross-sections for charged sleptons receive large corrections
near threshold from Coulombic photon exchange between the slowly moving
sleptons. Beamstrahlung and initial-state radiation also modify the excitation
curves substantially.


For slepton production in the continuum, {\it i.e.} sufficiently far above the
threshold, one can assume, to good approximation, on-shell production of the
slepton and thus factorize their decay. The cross-sections and polarization
asymmetries can be significantly modified by higher-order corrections. For all
production processes the most important decay modes, complete one-loop
corrections are available: in Ref.~\cite{slep} the $\OO(\alpha)$ corrections to
smuon, selectron and sneutrino production were presented, while stau
production, including mixing effects, has been analyzed in Ref.~\cite{stau}.
The slepton decays to leptons and neutralinos or
charginos were calculated at $\OO(\alpha)$ in Ref.~\cite{Guasch:01}.

\begin{figure}
\centering
\vspace*{5mm} 
\epsfig{figure=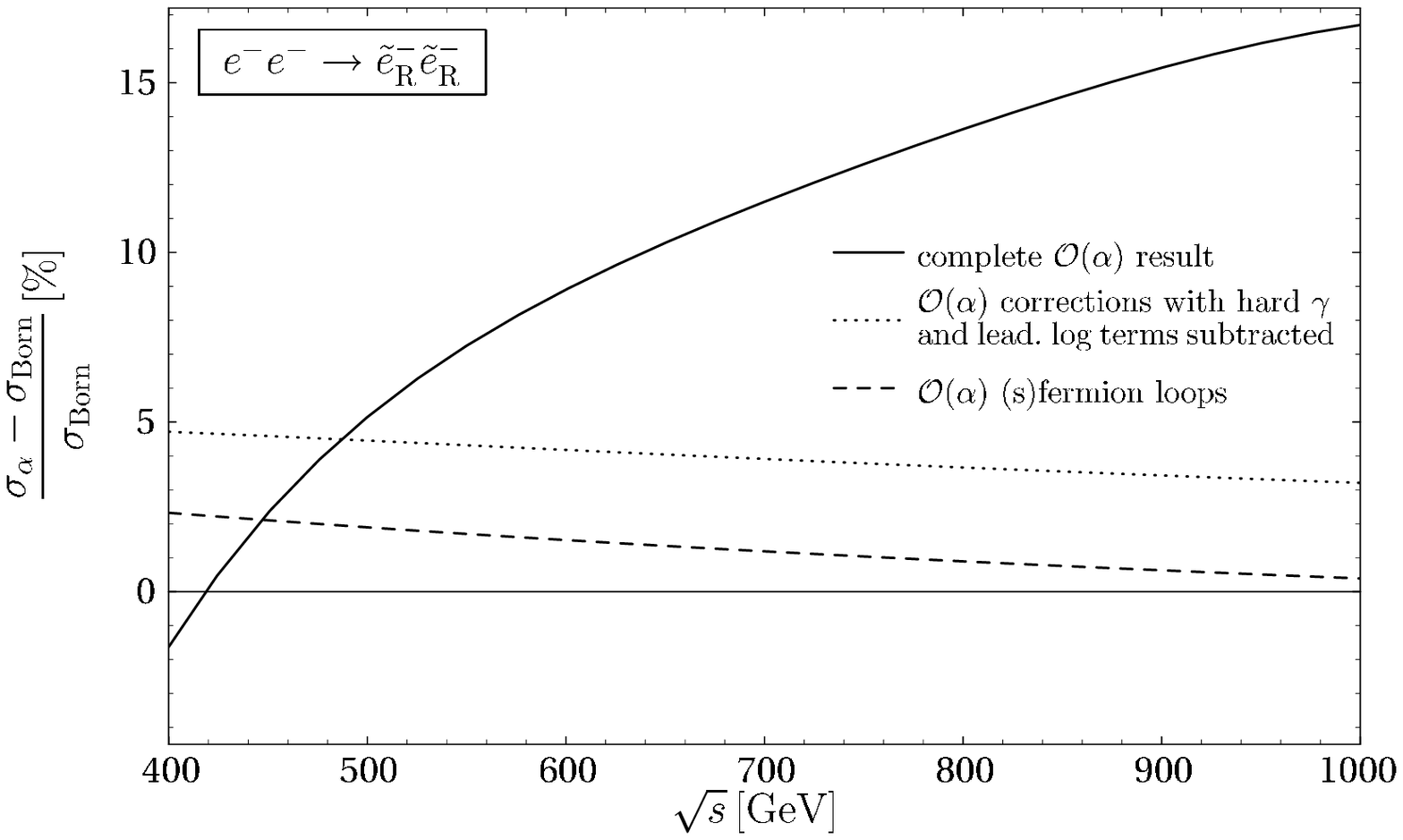, height=6cm, bb=-6 399 474 680}\\[1em]
\vspace*{5mm}
\epsfig{figure=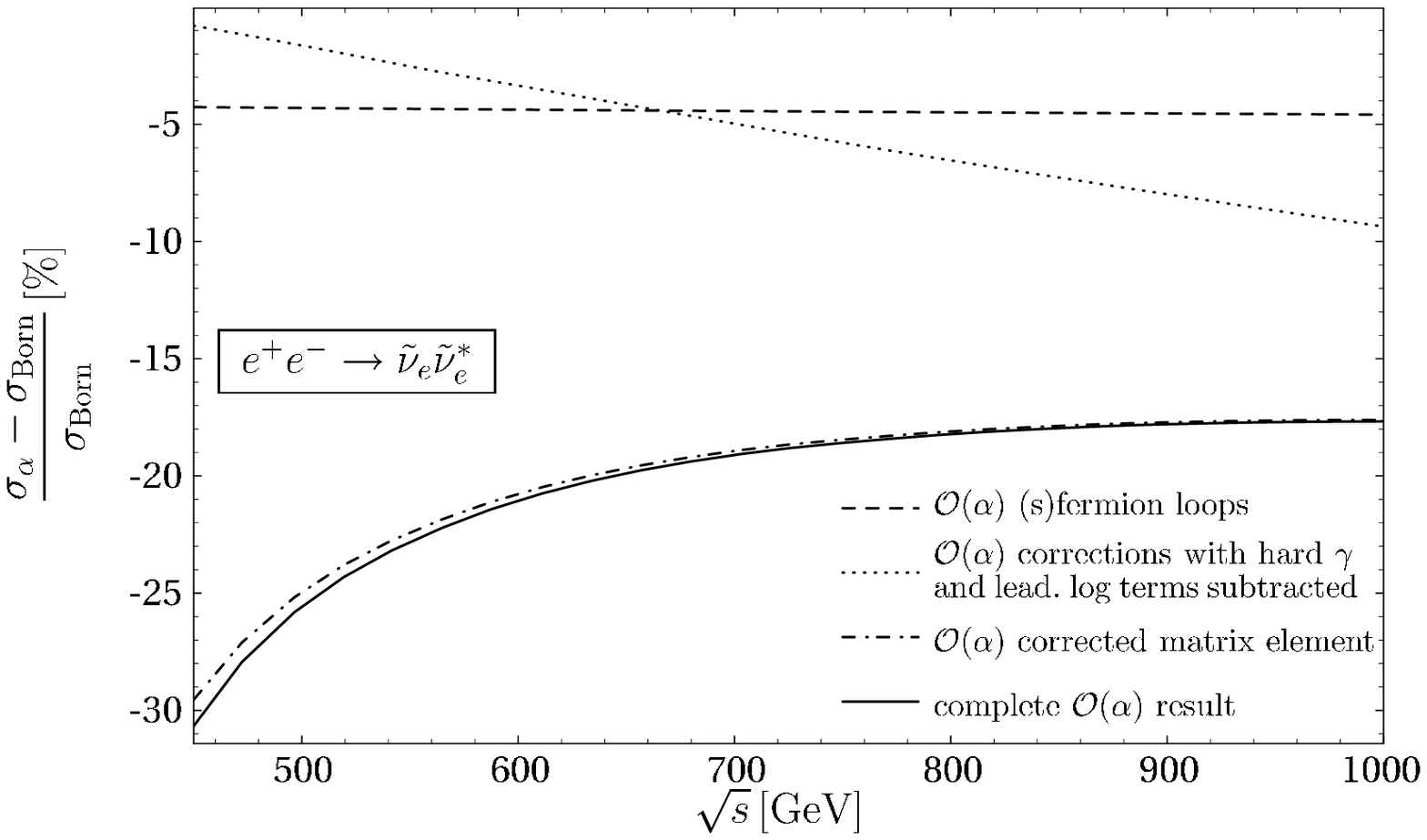, height=6cm}
\caption{$\OO(\alpha)$ corrections to the cross-sections
for $e^+e^- \to \se^-_R \se^-_R$ and
$e^+e^- \to \sne \sne^*$
relative to the Born
cross-sections. Besides the full $\OO(\alpha)$ result, contributions from
different subsets of diagrams are shown,
in particular the genuinely process-specific corrections 
defined by subtracting hard photon
radiation and leading-log soft and virtual photon effects from the overall $\OO(\alpha)$
corrections. The shift between the lower two curves in the $\sne$ panel
results from the one-loop
correction to the sneutrino mass.}
\label{fig:sne}
\end{figure}

For a full one-loop analysis, these individual calculations have to be
combined within a framework that uniquely defines the SUSY parameters beyond
tree-level. Such a comprehensive study is currently pursued within the SPA
Project \cite{spa,spatalks}. In the ``SPA Convention'' 
all heavy particle masses and
lepton masses are defined on-shell, while the MSSM Lagrangian parameters are 
given in the \drbar scheme at the scale $\tilde{M} = 1$ TeV.
This provides the basis from which all other parameters, {\it e.g.} the
neutralino/chargino masses and mixings, can be calculated at one-loop level in
any given scheme as part of the higher-order cross-section
and decay width calculations \cite{spatalk2}.
As an example, Fig.~\ref{fig:sne} shows the radiative corrections 
to 
selectron
pair production, $e^-e^- \to \se^-\se^-$ and
electron-sneutrino
pair production, $e^+e^- \to \sne\sne^*$, for the SPS1a scenario~\cite{sps}.

\section{Experimental Methods: Masses, Mixings, Couplings}

\begin{figure}
 (a) \\[-2ex]
\rule{0mm}{0mm} \hspace{1em}
\vspace*{10mm}
\epsfig{figure=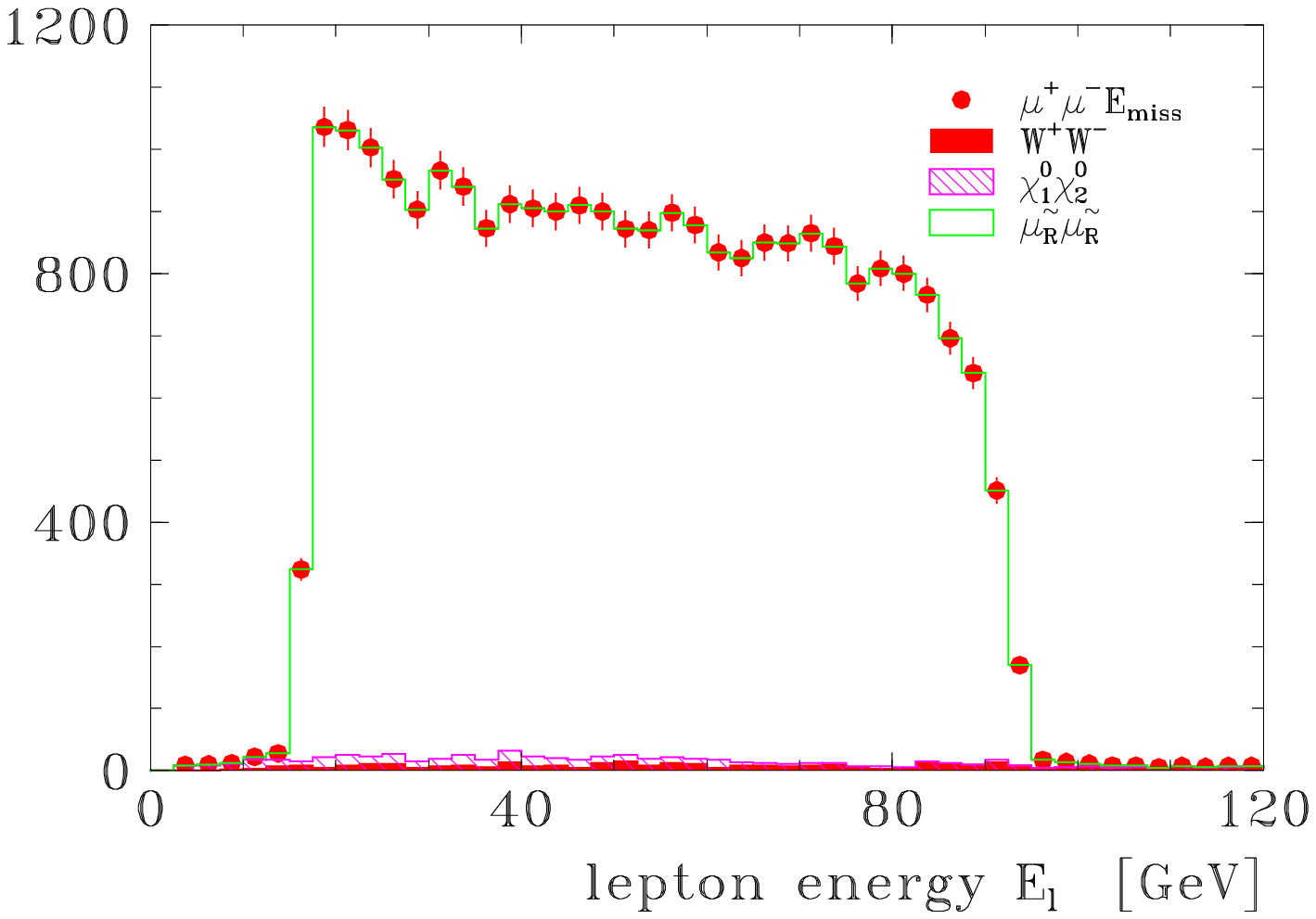, width=10cm} \\
 (b) \\[-2ex]
\rule{0mm}{0mm} \hspace{1.5em} 
\vspace*{10mm}
\psfig{figure=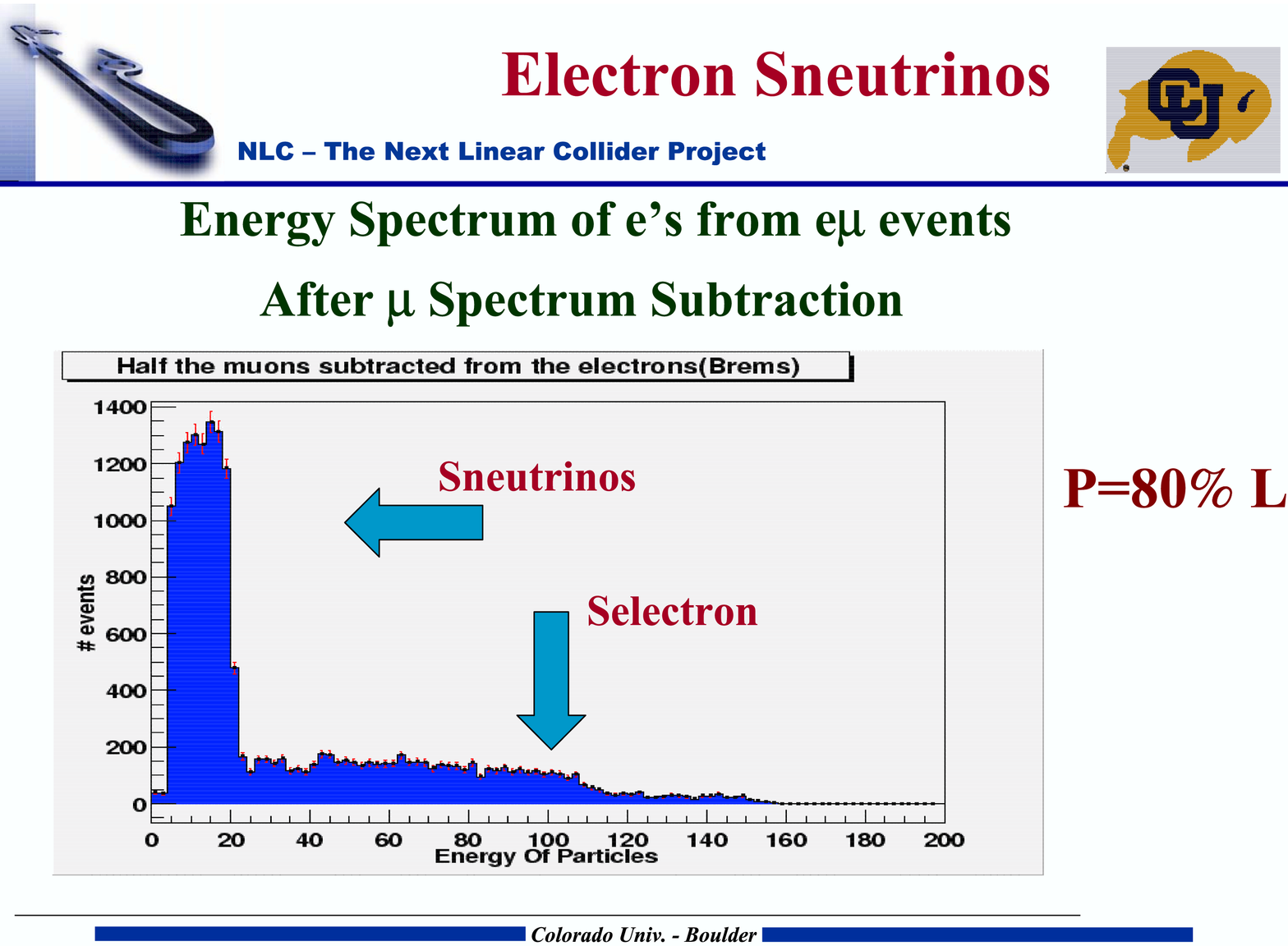, height=10cm, angle=270, bb=250 30 550 640, clip=true}
\caption{Charged
lepton energy spectra for (a) $e^+e^- \to \smuR^+\smuR^-$, $\smuR^\pm \to
\mu^\pm \neu_1$, and for (b) $e^+e^- \to \sne\sne^* \to \nu_e \neu_1 \, e^\pm
\cha^\mp_1 \to e^\pm \mu^\mp + /\!\!\!\!E$ including background selectron
production.}
\label{fig:spec}
\end{figure}

Based on the characteristic decay of sleptons into neutralinos, $\sll{} \to l
\, \neu_1$, both the masses of the slepton and neutralino can be determined
from measuring the upper and lower edges of the decay lepton spectrum
\cite{martyn}, see Fig.~\ref{fig:spec}~(a). The accuracy of such an analysis is
limited due to the high correlation between the slepton and neutralino mass
dependence.
The analysis of sneutrinos is more
involved for models with light sneutrinos, such as SPS1a, 
since in such scenarios most sneutrinos
decay into invisible channels \footnote{If the sneutrinos are
even lighter than the $\cha^\pm_1$ charginos, leading to completely invisible
sneutrino decays, the exciting opportunity opens up to study the sneutrino
properties through the lepton spectrum of the chargino decays \cite{lisn}.}, 
see Tab.~\ref{tab:sps1}.
The $\sn$ mass resolution could be optimized by focusing on the channel where one
sneutrino decays invisibly, while the other decays into a chargino,
$e^+e^- \to \sne\sne^* \to \nu_e \neu_1 \, e^\pm \cha^\mp_1$~\cite{nauenberg}, 
see Fig.~\ref{fig:spec}~(b).

Alternatively, the slepton masses can be determined from a scan of the
characteristic excitation curves for pair production near threshold,
leading in many channels to a precision superior to the decay
spectrum analysis \cite{slep}. In addition, one can use the extremely precise
determination of the R-selectron mass of $\Delta \mseR = 50$ MeV
from the threshold scan in $e^-e^-$ collisions
as an input for the analysis of the selectron decay energy spectrum, in order to
obtain a more accurate determination of the lightest neutralino mass. This
information is summarized in Tab.~\ref{tab:masses}. The numbers in parantheses
are estimates from Ref.~\cite{grannis:02}.
\begin{table}[tb]
\centering
\begin{tabular}{|c|c||c|c|c||c|}
\hline
               & $m$ & \multicolumn{3}{c||}{$\Delta m$ [GeV]} & $\Gamma$ \\
               & [GeV] & spectra & thr. scans & combined & [GeV] \\
\hline
$\tilde{\chi}^0_1$   &  96.1           & 0.10   & --   & 0.065 &
--
\\
\hline
$\tilde{e}_R$        & 143.0           & 0.08      & 0.05 & 0.05 &
        $0.21 \pm 0.05$ \\
$\tilde{e}_L$        & 202.1           & 0.8       & 0.2 & 0.2 &
 $0.25 \pm 0.02$ \\
$\tilde{\nu}_e$      & 186.0           & 1.2       & 1.1 & 1.1 &
 $<0.9$  \\
$\tilde{\mu}_R$      & 143.0           & 0.2       & 0.2 & 0.085
&
 $0.2 \pm 0.2$ \\
$\tilde{\mu}_L$      & 202.1           & --       &  (0.5) &&        \\
$\tilde{\tau}_1$     & 133.2           & 0.3      &        &&        \\
$\tilde{\tau}_2$     & 206.1           &          & (1.1)       &&   \\
\hline
\end{tabular}
\caption{Expected accuracies $\Delta m$ for slepton mass measurements via decay
energy spectra (3rd column) and threshold scans (4th col.) in the SPS1a
scenario. The combined values (5th col.) are based on using the threshold
R-selectron mass measurement as input for the decay spectrum analyses to
reduce correlation effects. The threshold scans are also sensitive to the 
slepton widths $\Gamma$ (6th col.).
The numbers in parentheses are estimates.
Void entries have not been analyzed yet; barred entries cannot be
accessed in the reference point SPS1a.}
\label{tab:masses}
\end{table}

\begin{figure}[t]
\psfig{figure=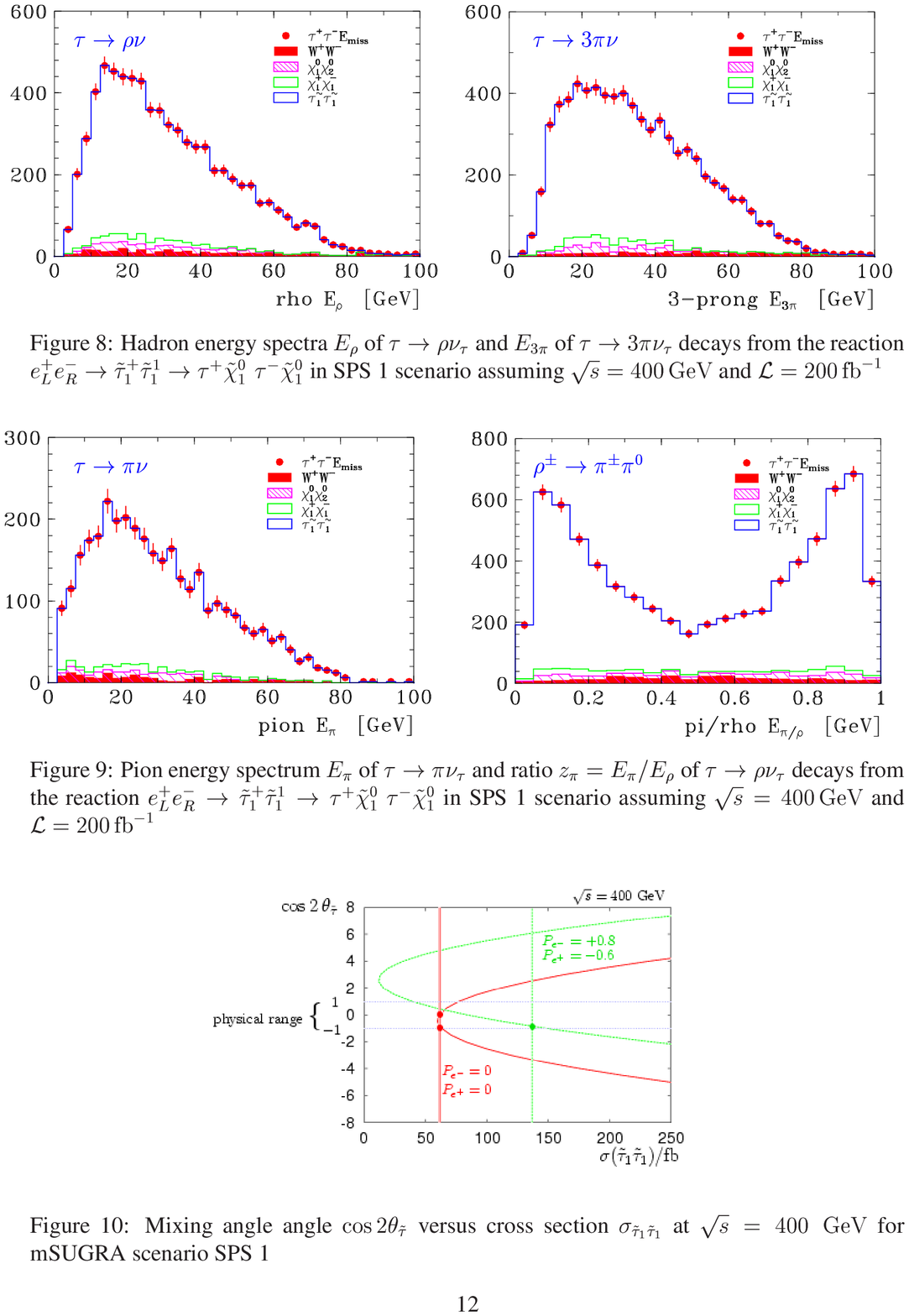, width=10cm,
        bb=165 155 420 300, clip=true}
\vspace{-1pt}
  \caption{\label{sigpol} {\it
      Mixing angle $\cos 2\theta_\st$ versus cross section
      $\sigma(e^+ e^- \to \st_1\st_1)$ at $\sqrt{s}=500~GeV$
      for beam polarisations $P_{e^-}=+0.8$ and  $P_{e^+}=-0.6$ (green/light)
      and the unpolarised case (red/dark).
      The vertical lines indicate the predicted cross sections.
}}
\end{figure}

Among the charged sleptons of the third generation, large mixing effects are
expected. While the stau masses can be determined using the same methods as
described above, the stau mixing angle $\theta_\st$ can be extracted from two
cross-section measurements $\sigma[e^+e^- \to \st_1 \st_1]$ with different beam
polarizations, see Fig.~\ref{sigpol}, Ref.~\cite{taupol}. 
In the SPS1a scenario one obtains $\cos 2\theta_\st = -0.84 \pm
0.04$, {\it i.e.} a precision is achievable at the per-cent level 
\cite{martyn}.
The value of the mixing angle and the degree of $\tau$ polarization in $\st_1$
decays depends on the fundamental parameters $\mu$, $A_\tau$ and $\tan\beta$ in
the Lagrangian, which can therefore be constrained by these measurements.
Since in a scenario with $\tan\beta \gesim 10$, charginos and neutralinos in the
decay chain will dominantly lead to additional tau leptons in the final state,
it is very difficult to disentangle the heavier $\st_2$ from the background of
the lighter $\st_1$. The $m_{\st_2}$ measurement is
therefore still an open problem.


The production cross-sections of the first generation sleptons are sensitive to
the SUSY Yukawa couplings between a selectron/sneutrino, electron and
neutralino/chargino, $\hat{g}(e\se\neu)$/$\hat{g}(e\sne\cha^\pm)$. From the
measurement of the $\seR$, $\seL$ and $\sne$ cross-sections one can therefore
test the fundamental SUSY identity between the gauge couplings $g$ and the
corresponding gaugino Yukawa couplings $\hat{g}$ in the electroweak sector. For
selectron production, beam polarization is crucial for disentangling 
the SU(2) and
U(1) Yukawa couplings. Taking into account uncertainties from the
selectron mass and the neutralino parameters, 
the SU(2) and U(1) Yukawa couplings,
$\hat{g}$ and $\hat{g}'$, can be extracted with a precision of 0.7\% and 0.2\%,
respectively, at a 500 GeV collider with 500 fb$^{-1}$ integrated luminosity
in the SPS1a scenario \cite{slep}.
Sneutrino production is only sensitive to the SU(2) coupling $\hat{g}$. Here the
dominantly invisible decay of the sneutrinos limits the expected precision, 
resulting in an error of 5\% \cite{slep}
in the SPS1a scenario.

\section{Conclusions}

The slepton sector is so far the best understood SUSY sector, 
both theoretically and experimentally. It has been shown in 
experimental simulation analyses that all
relevant slepton parameters can be measured 
at a linear collider
with per-cent or even per-mille
accuracy. The theoretical calculations for the cross-sections 
are under control only at the
per-cent level, so that much more work on the theoretical side is needed 
to match the
experimental accuracy. With all tools in place, the 
fundamental identity between gauge and Yukawa couplings can be
precisely tested and the accurate determination of the SUSY mass parameters and
mixings opens up a window to the understanding of the underlying 
fundamental supersymmetric theory and the
breaking of supersymmetry.

\end{document}